\documentclass[aps,prc,letterpaper,11pt,twoside,tightenlines,nofootinbib,showpacs,preprint,superscriptaddress]{revtex4}
\usepackage{amsmath}
\usepackage{graphicx}
\usepackage{subfigure}

\newcommand{\Ima}{\textrm{Im}}

\newcommand{\mev}{\textrm{ MeV}}

\begin{document}

\title{$J/\psi (\eta_c) N$ and $\Upsilon (\eta_b) N$ cross sections}

\author{C. W. Xiao}
\affiliation{Institut  f\"{u}r Kernphysik (Theorie), Institute for Advanced Simulation, and J\"ulich Center for Hadron Physics, Forschungszentrum J\"ulich, D-52425 J\"{u}lich, Germany}

\author{U.-G.~Mei{\ss}ner}
\affiliation{Helmholtz-Institut f\"ur Strahlen- und Kernphysik, and Bethe Center for Theoretical Physics, Universit\"at Bonn, D-53115  Bonn, Germany}
\affiliation{Institut  f\"{u}r Kernphysik (Theorie), Institute for Advanced Simulation, and J\"ulich Center for Hadron Physics, Forschungszentrum J\"ulich, D-52425 J\"{u}lich, Germany}

\begin{abstract}
Inspired by the recent findings of the two $P_c^+$ states in the $J/\psi p$ mass spectrum at LHCb, we investigate the elastic and inelastic cross sections of the $J/\psi N$, $\eta_c N$, $\Upsilon N$ and $\eta_b N$ channels within the constraints from  heavy quark spin and flavour symmetry. The $\bar{D}^{(*)} \Sigma_c^{(*)}$ ($B^{(*)} \Sigma_b^{(*)}$) bound states predicted  in earlier  works should be accessible  in elastic and/or inelastic processes of the $J/\psi N$ and/or $\eta_c N$ ($\Upsilon N$ and/or $\eta_b N$) interactions. 

\end{abstract}

\maketitle
\date{}

\section{Introduction}

Recently, the LHCb collaboration has reported the observation of two new $P_c^+$ states \cite{Aaij:2015tga}, $P_c(4380)^+, \, (\Gamma = 205\mev)$ (denoted as $P_1$) and $P_c(4450)^+, \, (\Gamma = 39\mev)$ ($P_2$), which are found in the $J/\psi p$ mass spectrum of the $\Lambda_b^0 \to J/\psi K^- p$ decays and consistent with pentaquark states. There are still some uncertainties in the LHCb analysis about the spin-parity $J^P$ quantum numbers,   with the possible assignments $(3/2^-, \, 5/2^+)$ respectively, or $(3/2^-, \, 5/2^+)$,  $(5/2^+, \, 3/2^-)$. What is nature of these two states? Soon after the experimental finding, they are considered as molecular states \cite{Chen:2015loa,Chen:2015moa,Roca:2015dva,He:2015cea}, but the molecular components are different among these works.  Using the one-pion exchange model, Ref. \cite{Chen:2015loa} suggests that $P_1$ is a $\bar{D}^* \Sigma_c$ molecular state with ($I = 1/2, \, J = 3/2$) and $P_2$ $\bar{D}^* \Sigma_c^*$ state with ($I = 1/2, \, J = 5/2$). And, $P_1$ is also interpreted as a $\bar{D}^* \Sigma_c$ state with ($I = 1/2, \, J^P = 3/2^-$) but $P_2$ as a $\bar{D}^* \Lambda_c$ and $\bar{D} \Sigma_c^*$ mixture ($I = 1/2, \, J = 5/2^+$) within the QCD sum rule frame \cite{Chen:2015moa}. The work of Ref. \cite{Roca:2015dva} based on the chiral unitary approach claims the $P_2$ to be a mixed state made of $\bar{D}^* \Sigma_c$ and $\bar{D}^* \Sigma_c^*$ with $I = 1/2, \, J^P = 3/2^-$ by the analysis of the $\Lambda_b$ decays \cite{Roca:2015tea,Feijoo:2015cca}. Furthermore, with the formalism of an effective theory, Ref. \cite{He:2015cea} explains $P_1$ and $P_2$ as the molecular candidates of $\bar{D} \Sigma_c^*$ and $\bar{D}^* \Sigma_c$, respectively. But, Ref. \cite{Mironov:2015ica} questions the explanation of the molecular state for these two new findings and suggest that they should be considered as multiquark configuration at the quark level. Following, these two states can be described as a five quark state of the diquark picture in the short range interaction at the quark level \cite{Maiani:2015vwa,Lebed:2015tna,Anisovich:2015cia}. On the contrary, it is suggested that they could be just a kinematical effect (cusp effect) \cite{Guo:2015umn,Liu:2015fea,Mikhasenko:2015vca} generated from the  triangle diagram singularity. Even though Refs.~\cite{Guo:2015umn,Liu:2015fea,Mikhasenko:2015vca} conclude that the peaks are not real states but the singularity effect of the triangle diagram which leads to the cusp of the decay amplitudes, the authors of \cite{Guo:2015umn} comment with caution that the conclusion of the kinematical effect can be distinguished with the future experimental measurement of the $\chi_{c1} p$ mass distribution in the process of $\Lambda_b^0 \to K^- \chi_{c1} p$. Later, the $\chi_{c1} p$ component, making of the $P_2$ state, is investigated in the work of \cite{Meissner:2015mza} utilizing the Weinberg compositeness condition \cite{Weinberg:1965zz} and an interesting analogy to the scalar meson $f_0(980)$ is drawn. Even if there are different opinions on the structure of these two new states, one should keep in mind that both of them are found in the $J/\psi p$ mass spectrum. This is the motivation of the present work to look into the $J/\psi N$ cross sections, where $N$ represents the nucleon.

Due to the heavy quark masses of $c, \ b$, and the couplings not known in the heavy charm and beauty sectors, which lead to the symmetry breaking in the heavy sector, with some assumptions and extrapolations for the interaction Lagrangian, one can get some insight into the heavy charm and beauty baryons \cite{Mizutani:2006vq,Haidenbauer:2010ch,He:2011jp,Yang:2011wz,Lu:2014ina}. Working on the heavy sector, one should take into account the heavy quark spin and flavour symmetry \cite{Isgur:1989vq,Neubert:1993mb,MW00}, which has been applied to the heavy meson-meson interactions \cite{Guo:2009id,Guo:2013sya} and heavy meson-baryon interactions \cite{GarciaRecio:2008dp,Gamermann:2010zz}. On the other hand, by assuming the SU(4) symmetry in charm sector, the works of \cite{Wu:2010jy,Wu:2010vk,Wu:2010rv} have predicted several narrow $N^*$ and $\Lambda^*$ states in the hidden charm and hidden beauty sectors. Furthermore, combining the heavy quark spin symmetry (HQSS) and the extended local hidden gauge formalism \cite{Bando:1987br,Meissner:1987ge,Harada:2003jx}, Refs.~\cite{Xiao:2013yca,Xiao:2013jla} obtain consistent results with the former predictions in Refs.~\cite{Wu:2010jy,Wu:2010vk,Wu:2010rv} but with extra predictions, summarized as a $\bar{D} \Sigma_c$ ($B \Sigma_b$) bound state with $(J=1/2, \, I=1/2)$, a $\bar{D} \Sigma_c^*$ ($B \Sigma_b^*$) state with $(J=3/2, \, I=1/2)$, a $I=1/2$ $\bar{D}^* \Sigma_c$ ($B^* \Sigma_b$) state degenerate in $(J=1/2, \, 3/2)$ and a $I=1/2$ $\bar{D}^* \Sigma_c^*$ ($B^* \Sigma_b^*$) state degenerate in $(J=1/2, \, 3/2, \, 5/2)$, where the results of hidden beauty baryons are analogous to the ones in the hidden charm sector because of the heavy quark flavour symmetry. Based on these two works, we study the cross sections of the $J/\psi N$ and $\eta_c N$ ($\Upsilon N$ and $\eta_b N$) channels in the present work. In the next section, our formalism is presented. Then, we show our results. Finally, we finish with the conclusions.

\section{Formalism}

Following the works of \cite{Xiao:2013yca,Xiao:2013jla}, we focus on the scattering amplitudes of the $J/\psi N$ and $\eta_c N$ ($\Upsilon N$ and $\eta_b N$) channels. The scattering amplitudes are evaluated by solving the coupled channels Bethe-Salpeter equation under the on shell factorization \cite{Oller:1997ti,Oset:1997it,Oller:2000fj}
\begin{equation}
T = [1 - V \, G]^{-1}\, V,
\label{eq:BS}
\end{equation}
where the propagator $G$ is a diagonal matrix with the loop functions of a meson and a baryon, the element of which is given by
\begin{equation}
G_{ii} (s) = i \int\frac{d^{4}q}{(2\pi)^{4}}\frac{2M_i}{(P-q)^{2}-M^2_i+i\varepsilon}\,\frac{1}{q^{2}-m^2_i+i\varepsilon},
\end{equation}
where $m_i, ~M_i$ are the masses of meson and baryon in $i^{\rm th}$ channel, respectively, $q$ is the four-momentum of the meson, and $P$ is the total four-momentum of the meson and the baryon, thus, $s=P^2$. The kernel matrix  $V$ contains the interaction potentials which are derived from the Lagrangian. In the present work, following 
Refs.~\cite{Xiao:2013yca,Xiao:2013jla}, we take the constraints from the HQSS into account, thus, the elements of $V_{ij}$ for the $(J=1/2, \, I=1/2)$ sector are given in 
Table~\ref{tab:vij11}, and for the $(J=3/2, \, I=1/2)$ sector in Table \ref{tab:vij31}, which are extrapolated to the hidden beauty sectors just by replacing the quark $\bar{c} \to \bar{b}$ for the corresponding mesons and $c \to b$ for the baryons.  In Tables~\ref{tab:vij11} and \ref{tab:vij31}, the coefficients $\mu_{i}^I$, $\mu_{ij}^I$ ($i,j=1,2,3$) and $\lambda_2^I$ are the unknown low energy constants under the HQSS bases, which specify the isospin sector and can be related using $SU(3)$ flavour symmetry. Note that all of them just depend on the isospin ($I$) sector and are independent of the spin $J$, which is a consequence of the HQSS constraints. The values of the coefficients are dependent on the model used. As discussed in the introduction, using the local hidden gauge formalism, we obtain their values for the two spin sectors
\begin{equation}
\begin{split}
\mu_2 &= \frac{1}{4f^2_\pi} (k^0 + k'^0),\\
\mu_3 &= -\frac{1}{4f^2_\pi} (k^0 + k'^0),\\
\mu_{12} &= -\sqrt{6}\ \frac{m_\rho^2}{p^2_{D^*} - m^2_{D^*}}\ \frac{1}{4f^2_\pi}\ (k^0 + k'^0),\\
\mu_1 &= 0,\qquad \mu_{23} = 0, \\
\lambda_2 &= \mu_3,\qquad \mu_{13} = -\mu_{12}~,
\end{split}
\end{equation}
where $p_{D^*}$ and $m_{D^*}$ are the four momentum and the mass of $D^*$ (for the hidden beauty cases just changing to the ones of $B^*$), $f_\pi$ the pion decay constant, and $k^0, k'^0$ are the energies of the incoming and outgoing mesons (for the vector mesons, we have ignored the factor $\vec{\epsilon} ~\vec{\epsilon}~'$).
Of course, these are only the lowest order interactions and eventually higher order terms need to be taken into account. For a first estimate has given here, such an
approach is, however, justified.

\begin{table}[ht]
     \renewcommand{\arraystretch}{1.7}
     \setlength{\tabcolsep}{0.4cm}
\centering
\caption{The elements  $V_{ij}$ corresponding to the channels in the  $J=1/2,~I=1/2$ sector.}
\label{tab:vij11}
\begin{tabular}{ccccccc}
\hline\hline
$\eta_c N$ & $J/\psi N$ &  $\bar D \Lambda_c$ &  $\bar D \Sigma_c$ &  $\bar D^* \Lambda_c$
  &  $\bar D^* \Sigma_c$ &  $\bar D^* \Sigma^*_c$   \\
\hline
$\mu_1$ & 0 & $\frac{\mu_{12}}{2}$ &
 $\frac{\mu_{13}}{2}$ & $\frac{\sqrt{3} \mu_{12}}{2}$ &
 $-\frac{\mu_{13}}{2 \sqrt{3}}$ & $\sqrt{\frac{2}{3}} \mu_{13}$ \\
  & $\mu_1$ & $\frac{\sqrt{3} \mu_{12}}{2}$ & $-\frac{\mu_{13}}{2 \sqrt{3}}$ & $-\frac{\mu_{12}}{2}$
 & $\frac{5 \mu_{13}}{6}$ & $\frac{\sqrt{2}\mu_{13}}{3}$ \\
  &  & $\mu_2$ & 0 & 0 & $\frac{\mu_{23}}{\sqrt{3}}$ & $\sqrt{\frac{2}{3}} \mu_{23}$ \\ 
  &  &  & $\frac{1}{3} (2 \lambda_2 + \mu_3)$ & $\frac{\mu_{23}}{\sqrt{3}}$ & $\frac{2 (\lambda_2 - \mu_3)}{3 \sqrt{3}}$ & $\frac{1}{3} \sqrt{\frac{2}{3}} (\mu_3-\lambda_2 )$ \\ 
  &  &  &  & $\mu_2$ & $-\frac{2 \mu_{23}}{3}$ & $\frac{\sqrt{2} \mu_{23}}{3}$ \\ 
  &  &  &  &  & $\frac{1}{9} (2 \lambda_2 +7 \mu_3)$ & $\frac{1}{9} \sqrt{2} (\mu_3-\lambda_2)$ \\ 
  &  &  &  &  &  & $\frac{1}{9} (\lambda_2+8 \mu_3)$ \\
\hline
\end{tabular}
\end{table}

\begin{table}[ht]
     \renewcommand{\arraystretch}{1.7}
     \setlength{\tabcolsep}{0.4cm}
\centering
\caption{The elements  $V_{ij}$ corresponding to  the channels in the $J=3/2,~I=1/2$ sector.}
\label{tab:vij31}
\begin{tabular}{ccccc}
\hline\hline
$J/\psi N$ &  $\bar D^* \Lambda_c$ &  $\bar D^* \Sigma_c$ 
&  $\bar D \Sigma^*_c$  &  $\bar D^* \Sigma^*_c$   \\
\hline
$\mu_1$ & $\mu_{12}$ & $\frac{\mu_{13}}{3}$ & $-\frac{\mu_{13}}{\sqrt{3}}$ & $\frac{\sqrt{5} \mu_{13}}{3}$ \\
  & $\mu_2$ & $\frac{\mu_{23}}{3}$ & $-\frac{\mu_{23}}{\sqrt{3}}$ & $\frac{\sqrt{5} \mu_{23}}{3}$ \\
  &  & $\frac{1}{9} (8 \lambda_2 + \mu_3)$ & $\frac{\lambda_2 - \mu_3}{3 \sqrt{3}}$ & $\frac{1}{9} \sqrt{5} (\mu_3-\lambda_2)$  \\ 
  &  &  & $\frac{1}{3} (2 \lambda_2 +\mu_3)$ &
   $\frac{1}{3} \sqrt{\frac{5}{3}} (\lambda_2 -\mu_3)$  \\ 
  &  &  &  & $\frac{1}{9} (4 \lambda_2 +5 \mu_3)$  \\ 
\hline
\end{tabular}
\end{table}

In the present work, we investigate the $J/\psi N$ and $\eta_c N$ ($\Upsilon N$ and $\eta_b N$) channels by evaluating their cross sections. Using the optical theorem, we obtain
\begin{equation}
\sigma_{tot} = -\frac{M_N}{P_{cm}^{J/\psi} \sqrt{s}} \Ima\;T_{J/\psi N \to J/\psi N},
\label{eq:sigtot}
\end{equation} 
where $P_{cm}^{J/\psi}$ is the momentum of $J/\psi$ in the center mass frame and $M_N$ the mass of nucleon, and one defines the elastic cross section
\begin{equation}
\sigma_{el} = \frac{1}{4 \pi} \frac{M_N^2}{s} \overline{\sum} \sum |T_{J/\psi N \to J/\psi N}|^2,
\label{eq:sigela}
\end{equation}
where $\sum, \, \overline{\sum}$ stand for sum and average over the spins of the nucleons and $J/\psi$. Hence, the inelastic cross section is given by
\begin{eqnarray}
\sigma_{in} &&= \sigma_{tot} - \sigma_{el} \nonumber \\
&&=-\frac{M_N}{P_{cm}^{J/\psi} \sqrt{s}} \Ima\;T_{J/\psi N \to J/\psi N} - \frac{1}{4 \pi} \frac{M_N^2}{s} \overline{\sum} \sum |T_{J/\psi N \to J/\psi N}|^2.\label{eq:sigine}
\end{eqnarray}

\section{Results}

\begin{figure}
\centering
\includegraphics[scale=0.42]{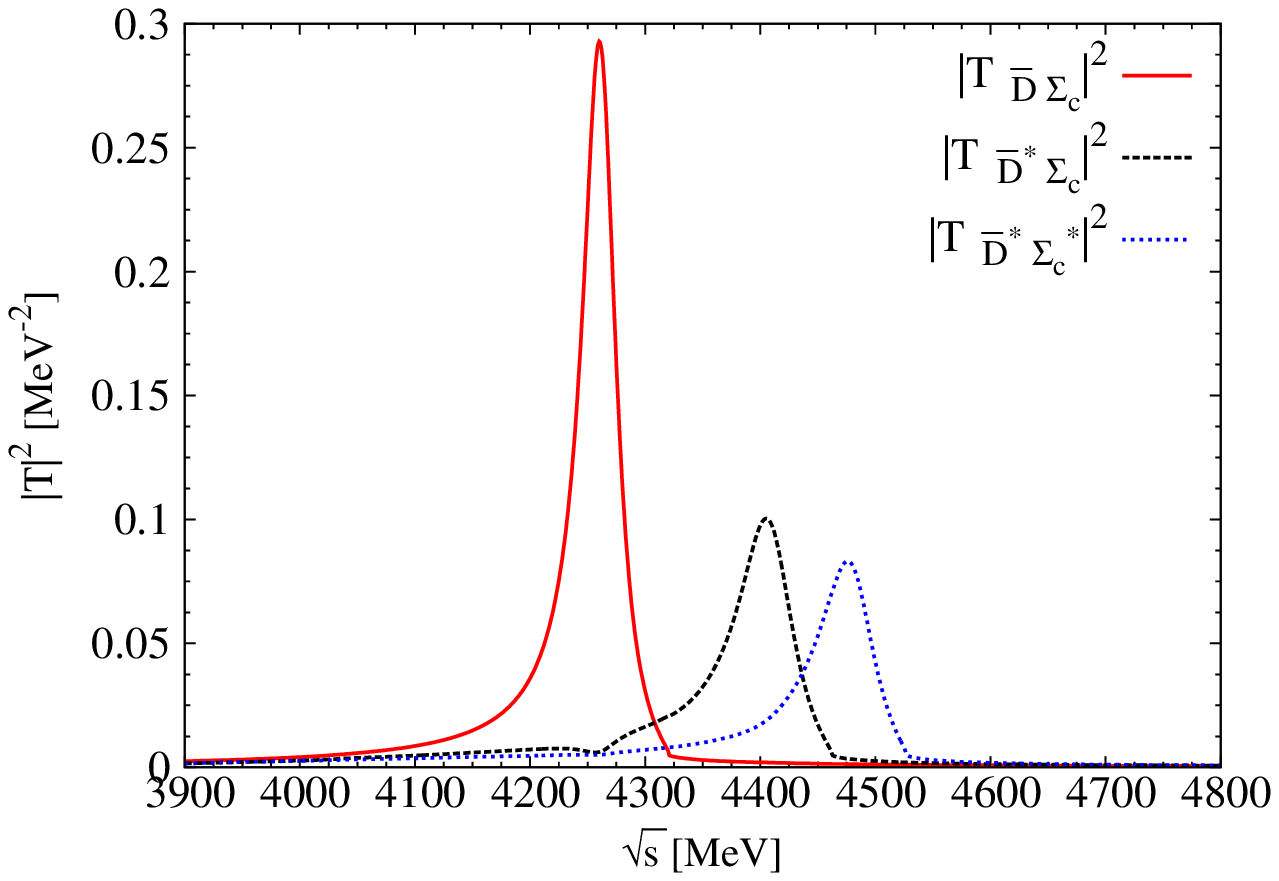}
\includegraphics[scale=0.42]{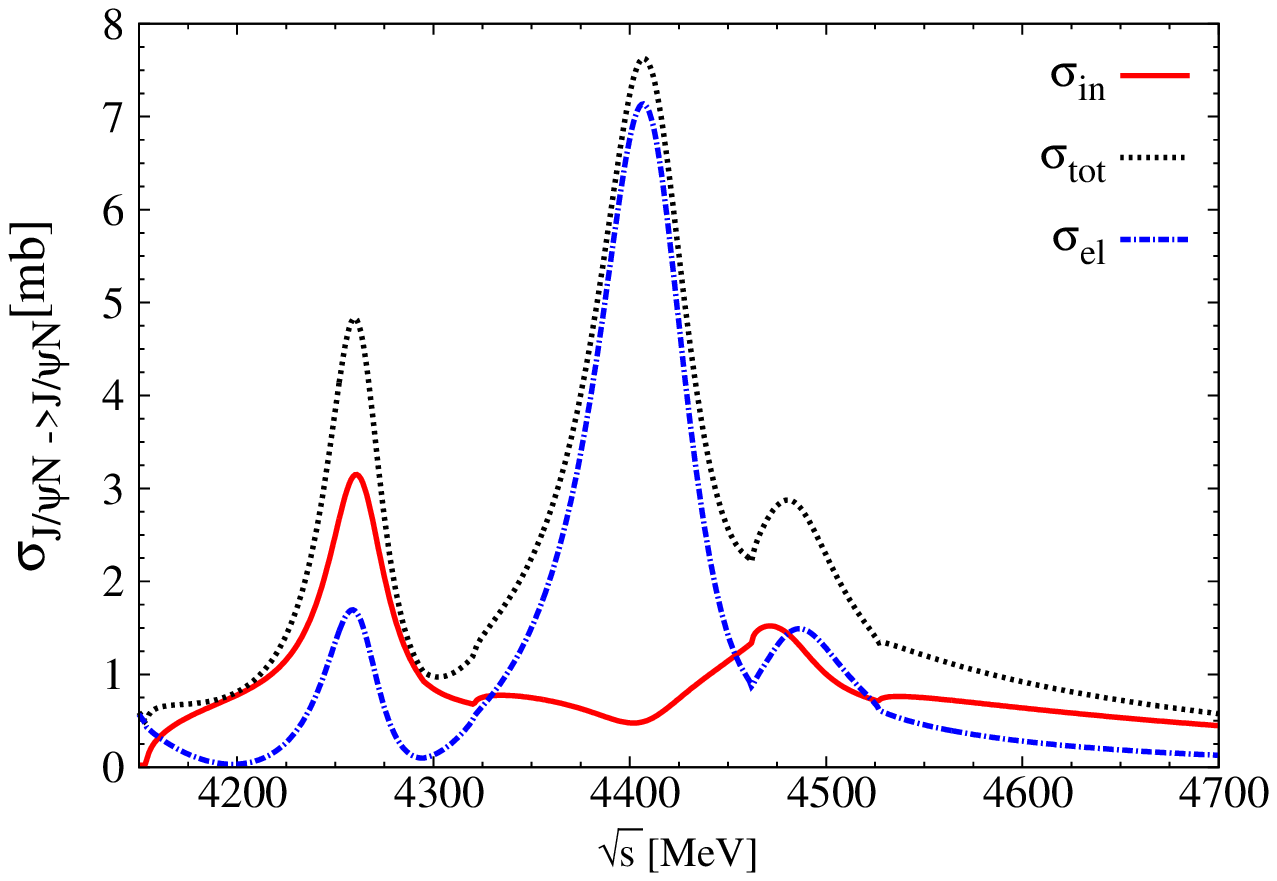}
\includegraphics[scale=0.42]{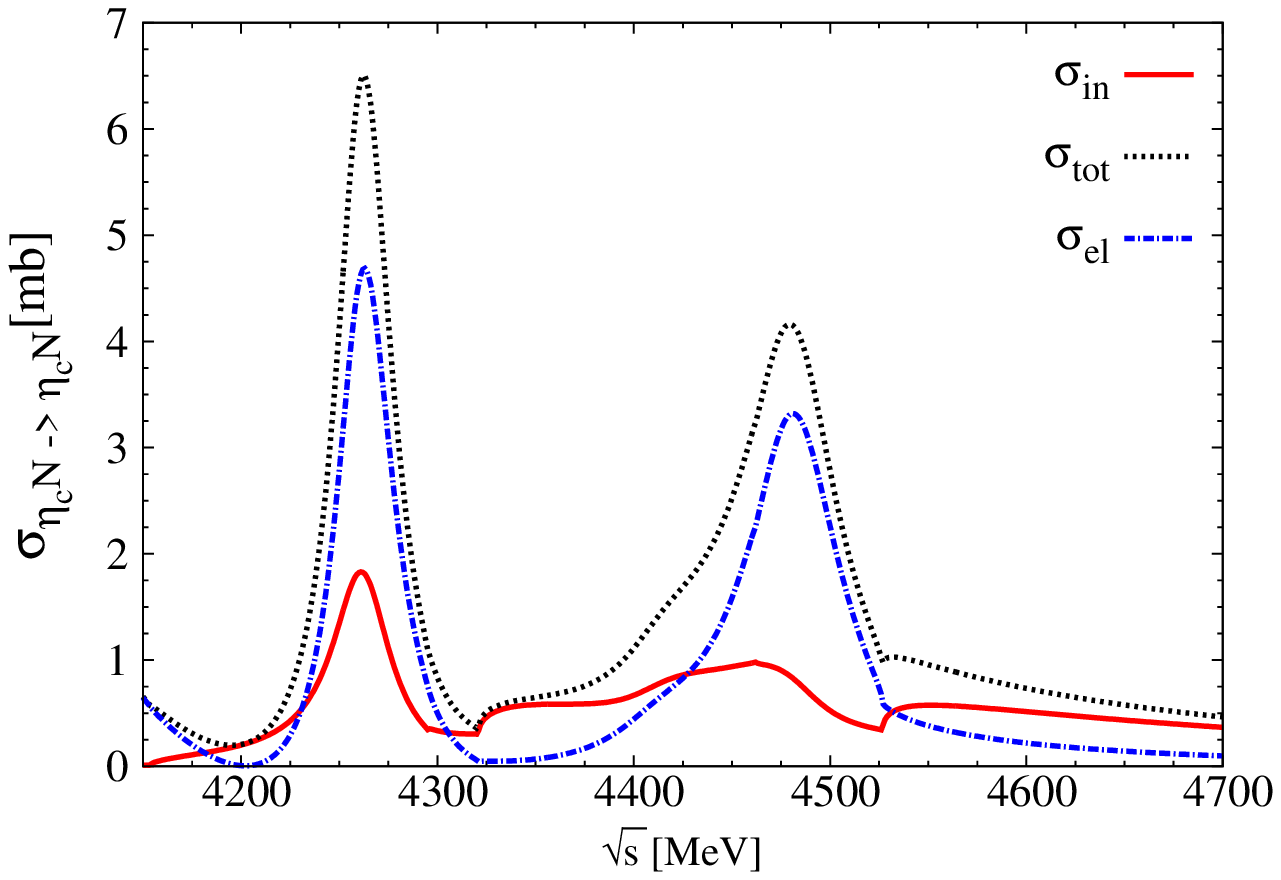}
\caption{Results of the $J=1/2,~I=1/2$ sector with hidden charm. Left: Modulus squared of the amplitudes. Middle: The total (dash/black line), elastic (dot-dash/blue line) and inelastic (solid/red line) cross sections of the $J/\psi N$ channel. Right: The total, elastic and inelastic cross sections for the $\eta_c N$ channel.}
\label{fig:charm1}
\end{figure}

Using the coupled channel approach, all the channels that we are concerned with can be seen in Tables~\ref{tab:vij11} and \ref{tab:vij31}, are $\eta_c N$, $J/\psi N$, $\bar D \Lambda_c$, $\bar D \Sigma_c$, $\bar D^* \Lambda_c$, $\bar D^* \Sigma_c$, $\bar D \Sigma_c^*$, $\bar D^* \Sigma^*_c$ for the hidden charm sector ($J=1/2, \ 3/2$), and $\eta_b N$, $\Upsilon N$, $B \Lambda_b$, $B \Sigma_b$, $B^* \Lambda_b$, $B^* \Sigma_b$, $B \Sigma_b^*$, $B^* \Sigma^*_b$ for the hidden beauty sector ($J=1/2, \ 3/2$). Then, with Eqs. \eqref{eq:BS}, \eqref{eq:sigtot}, \eqref{eq:sigela} and \eqref{eq:sigine}, we can evaluate the scattering amplitudes and the cross sections. We show our result of hidden charm in the $J=1/2,~I=1/2$ sector in Fig.~\ref{fig:charm1}. In the left of Fig.~\ref{fig:charm1}, we have reproduced the results of \cite{Xiao:2013yca}, where the three predicted bound states are clearly seen in the squares of the scattering amplitudes for the channels $\bar D \Sigma_c$, $\bar D^* \Sigma_c$, $\bar D^* \Sigma^*_c$ respectively,  with the masses $4262\mev, ~4410\mev, ~4481\mev$ and the widths $35\mev, ~58\mev, ~56\mev$ respectively. These three states are also seen in the total and elastic cross sections of the $J/\psi N$ channel, seen in the middle of Fig. \ref{fig:charm1}. Interestingly, the second state of $\bar D^* \Sigma_c$ can not be found in the inelastic cross section, which means that this state would not be seen in the inelastic process and coincides with the fact of the experimental finding for only two $P_c^+$ states found in the $J/\psi p$ mass spectrum in Ref.~\cite{Aaij:2015tga}. Note that our results do not take into account any background and the theoretical uncertainties. Even though we find three states in the two-body scattering, when we look into the inelastic scattering cross section, only two of them can be found. The results of Ref.~\cite{Molina:2012mv} reveal that the peak seen in the inelastic cross section would be destroyed by the Fermi motion of the nuclear target. But, these peaks can also be looked for in other inelastic processes, like decay model of Ref. \cite{Aaij:2015tga}. Thus, we suggest that our predicted states, $\bar D \Sigma_c$ and $\bar D^* \Sigma^*_c$, could be looked for in other inelastic processes of the $J/\psi N$ final state. When we look at the cross sections of the $\eta_c N$ channel in the right of Fig.~\ref{fig:charm1}, the state of $\bar D^* \Sigma_c$ is still missing in the total and elastic cross sections since it couples to $\eta_c N$ channel weakly \cite{Xiao:2013yca}, and the state of $\bar D^* \Sigma^*_c$ disappears in the inelastic cross section. To summarize, we find that only one of the $\bar D \Sigma_c$ is  found in both the inelastic scattering of the $J/\psi N$ and $\eta_c N$ final state interaction, the second one just can be seen in the the elastic scattering of the $J/\psi N$ interaction, and the one of $\bar D^* \Sigma^*_c$ can be looked for in the inelastic scattering of the $J/\psi N$ interaction and elastic scattering of the $\eta_c N$ channel.

\begin{figure}
\centering
\includegraphics[scale=0.42]{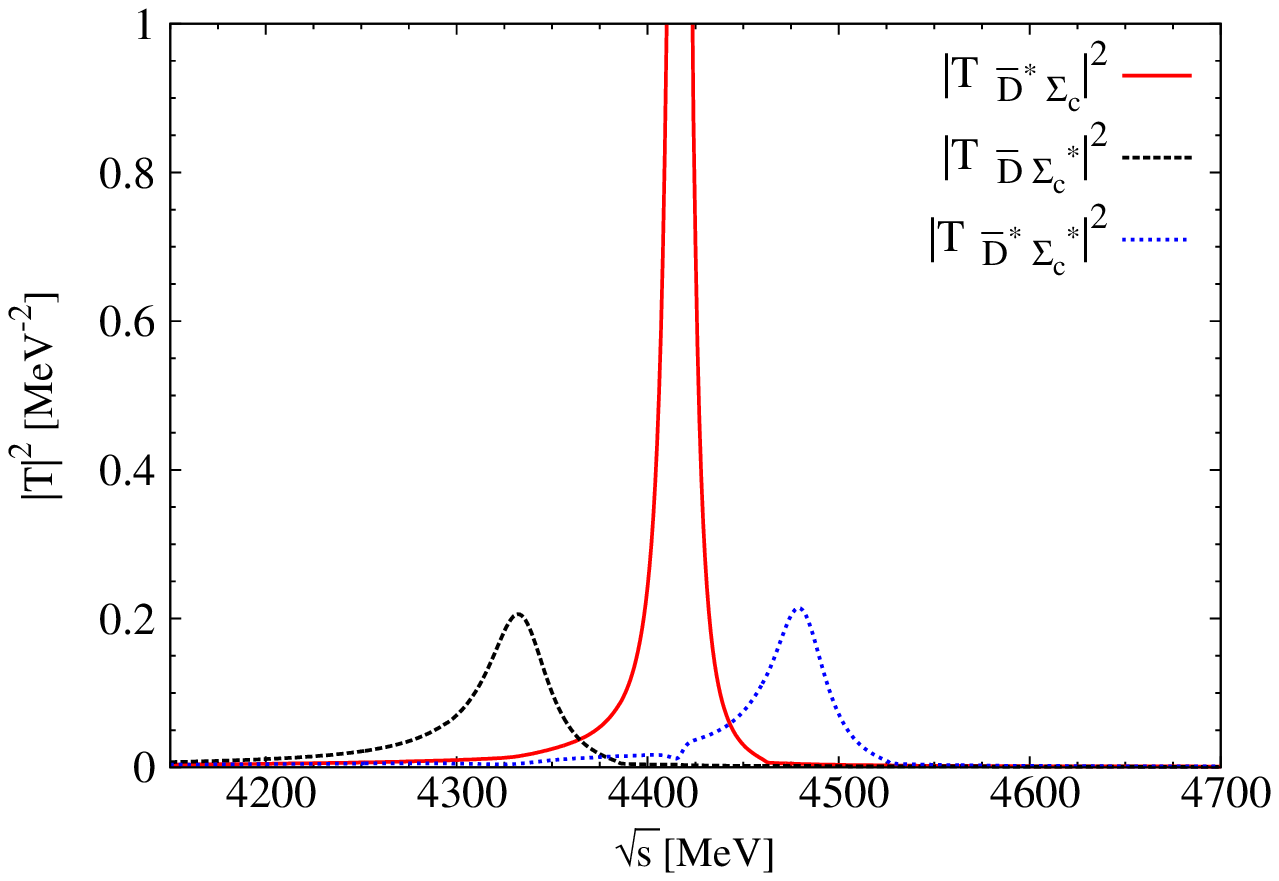}
\includegraphics[scale=0.42]{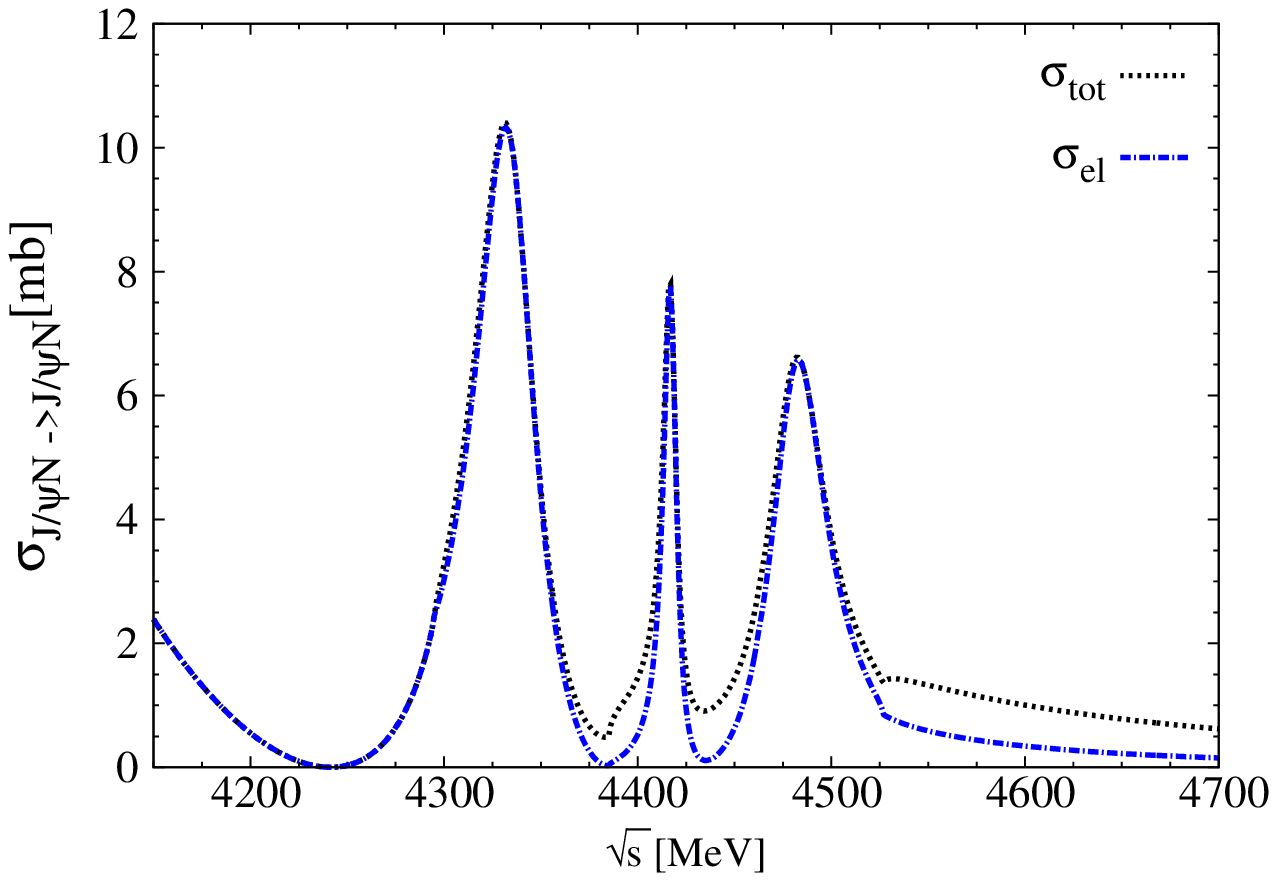}
\caption{Result of the $J=3/2,~I=1/2$ sector with hidden charm. Left: Modulus squared of the amplitudes. Right: The total (dash/black line) and elastic (dot-dash/blue line) cross sections of the $J/\psi N$ channel.}
\label{fig:charm2}
\end{figure}

The results of hidden charm in the $J=3/2,~I=1/2$ sector are shown in Fig.~\ref{fig:charm2}. The three predicted states \cite{Xiao:2013yca} are seen in the clear peaks of the modulus squared of the amplitudes in the left panel of Fig.~\ref{fig:charm2}, which are a bit below the thresholds of $\bar{D} \Sigma_c^*, ~\bar{D}^* \Sigma_c, ~\bar{D}^* \Sigma_c^*$ channels, respectively, with the masses $4334\mev, ~4417\mev, ~4481\mev$ and the widths $38\mev, ~8\mev, ~35\mev$ respectively. The structures of the three peaks corresponding to these states are also seen in the total and elastic cross sections of the $J/\psi N$ channel in the right panel of Fig.~\ref{fig:charm2}, where we find that the total cross section is essentially given by the elastic one. Therefore, these three states could  only be found in the elastic processes of the $J/\psi N$ interaction.

\begin{figure}
\centering
\includegraphics[scale=0.42]{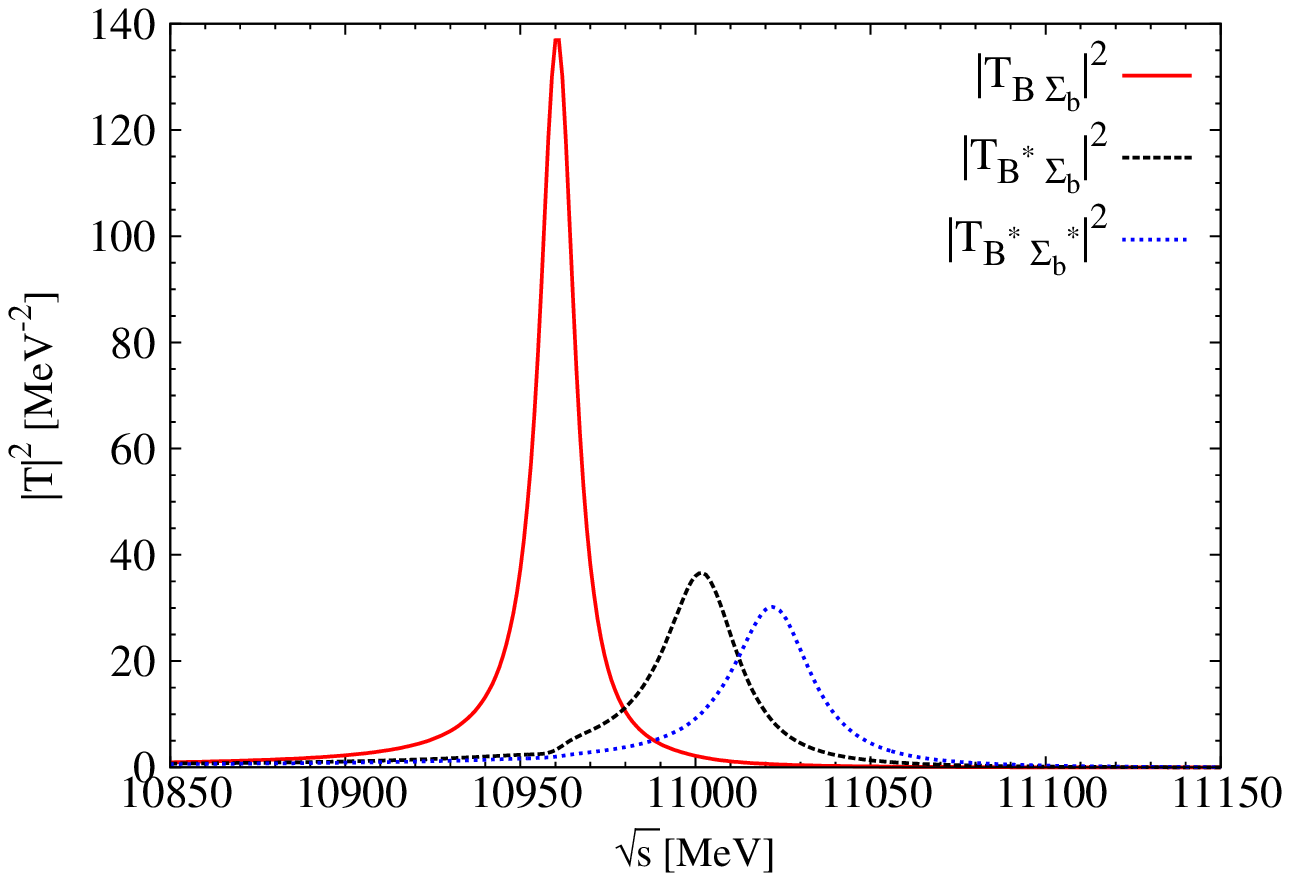}
\includegraphics[scale=0.42]{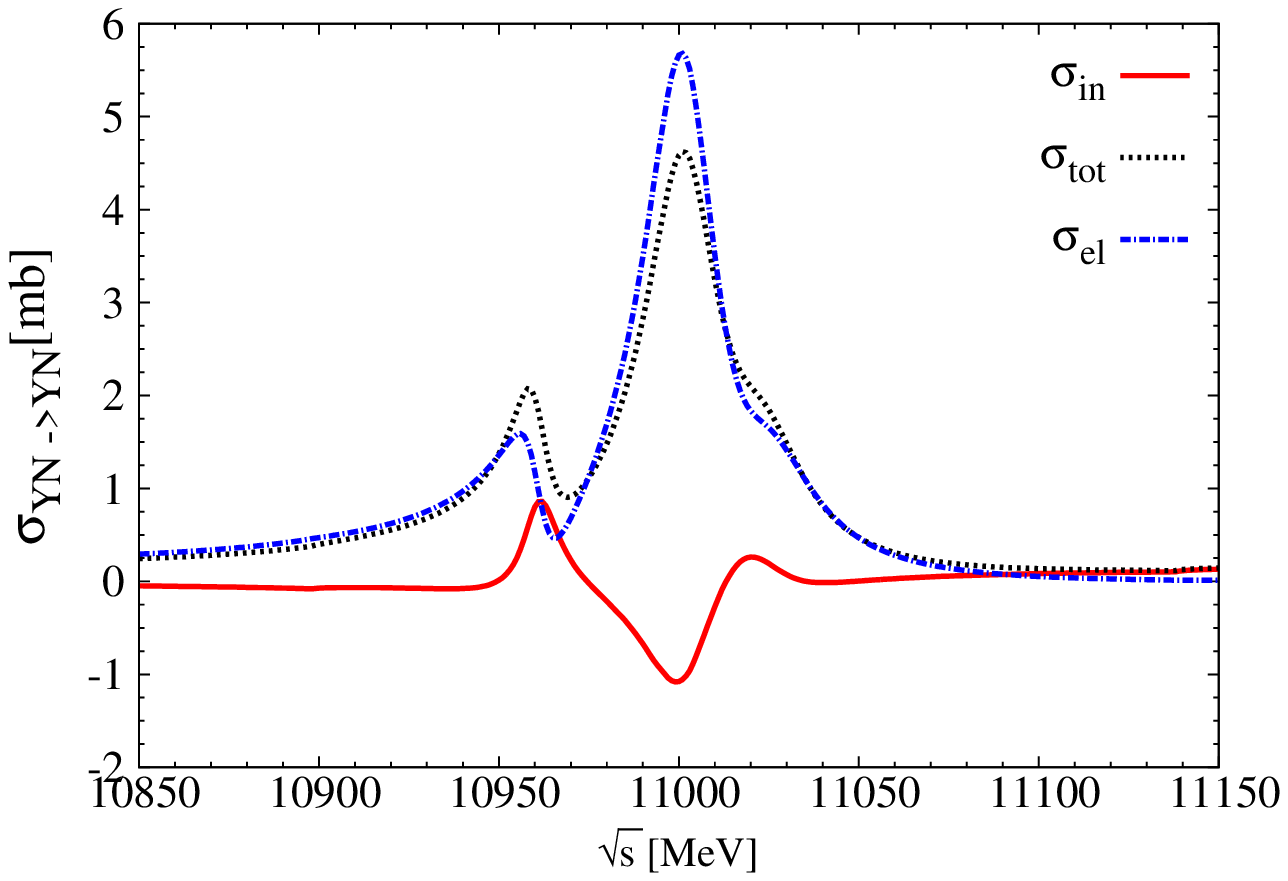}
\includegraphics[scale=0.42]{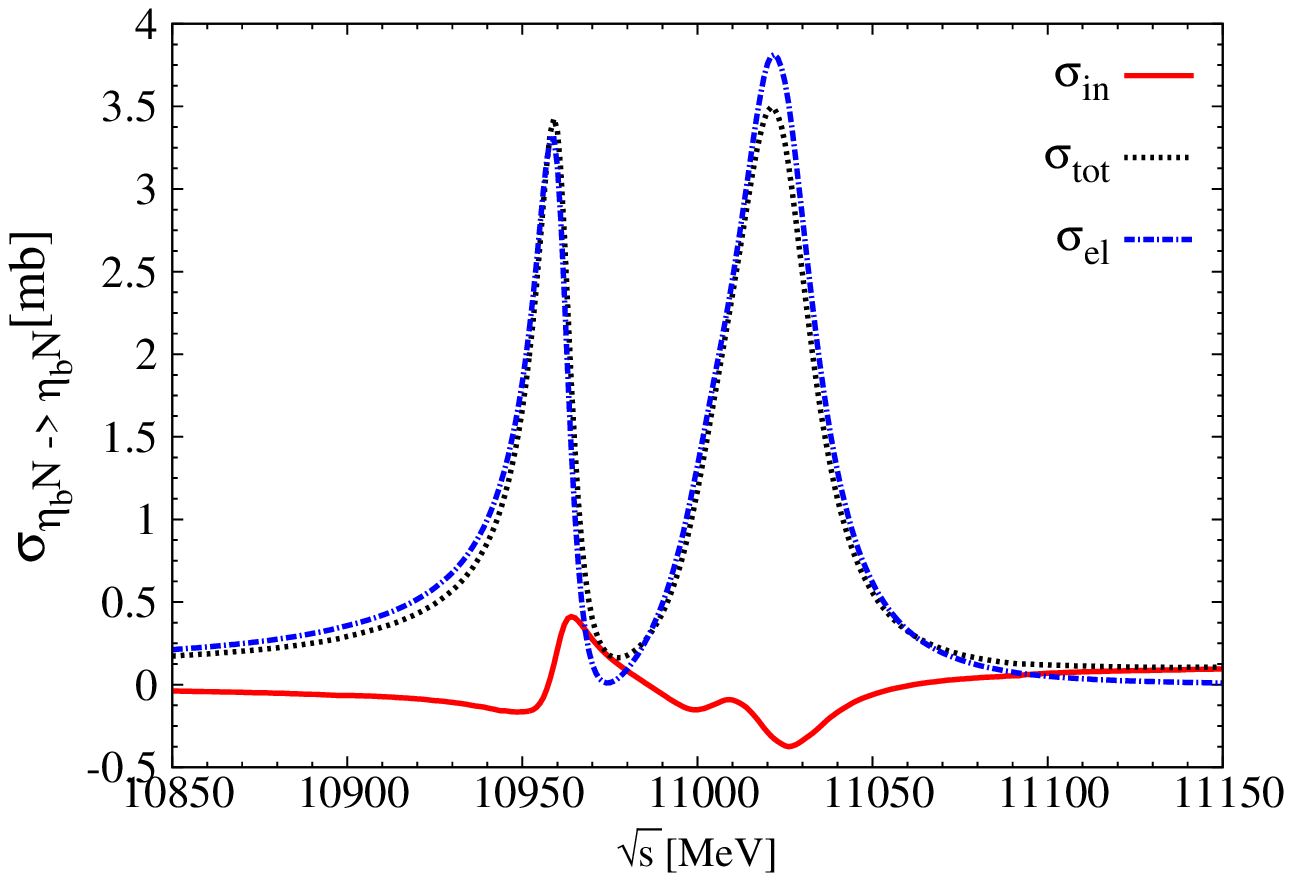}
\caption{Result of the $J=1/2,~I=1/2$ sector with hidden beauty. Left: Modulus squared of the amplitudes. Middle: The total, elastic and inelastic cross sections of the $\Upsilon N$ channel. Right: The total, elastic and inelastic cross sections for the $\eta_b N$ channel. (Analogous to Fig. \ref{fig:charm1})}
\label{fig:beauty1}
\end{figure}

\begin{figure}
\centering
\includegraphics[scale=0.42]{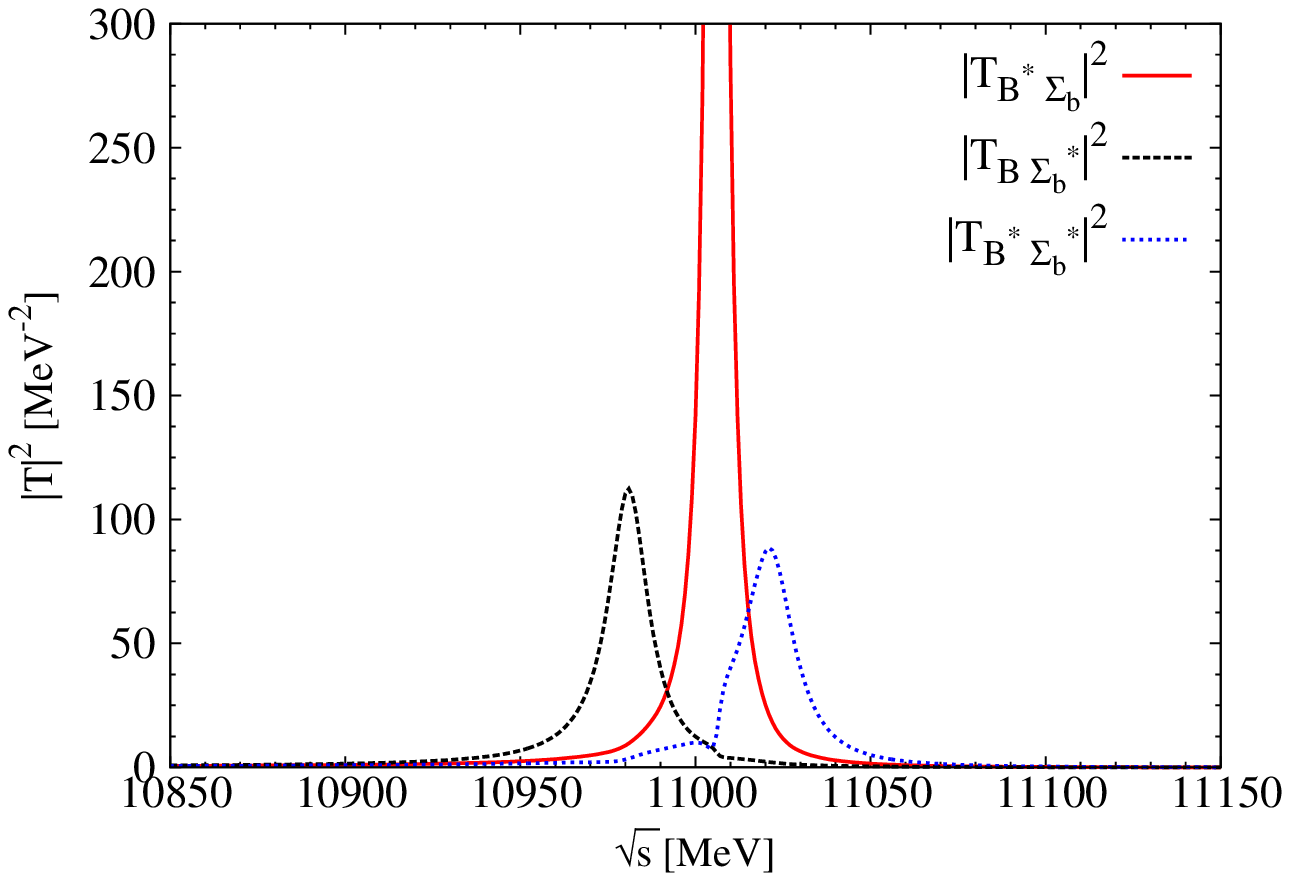}
\includegraphics[scale=0.42]{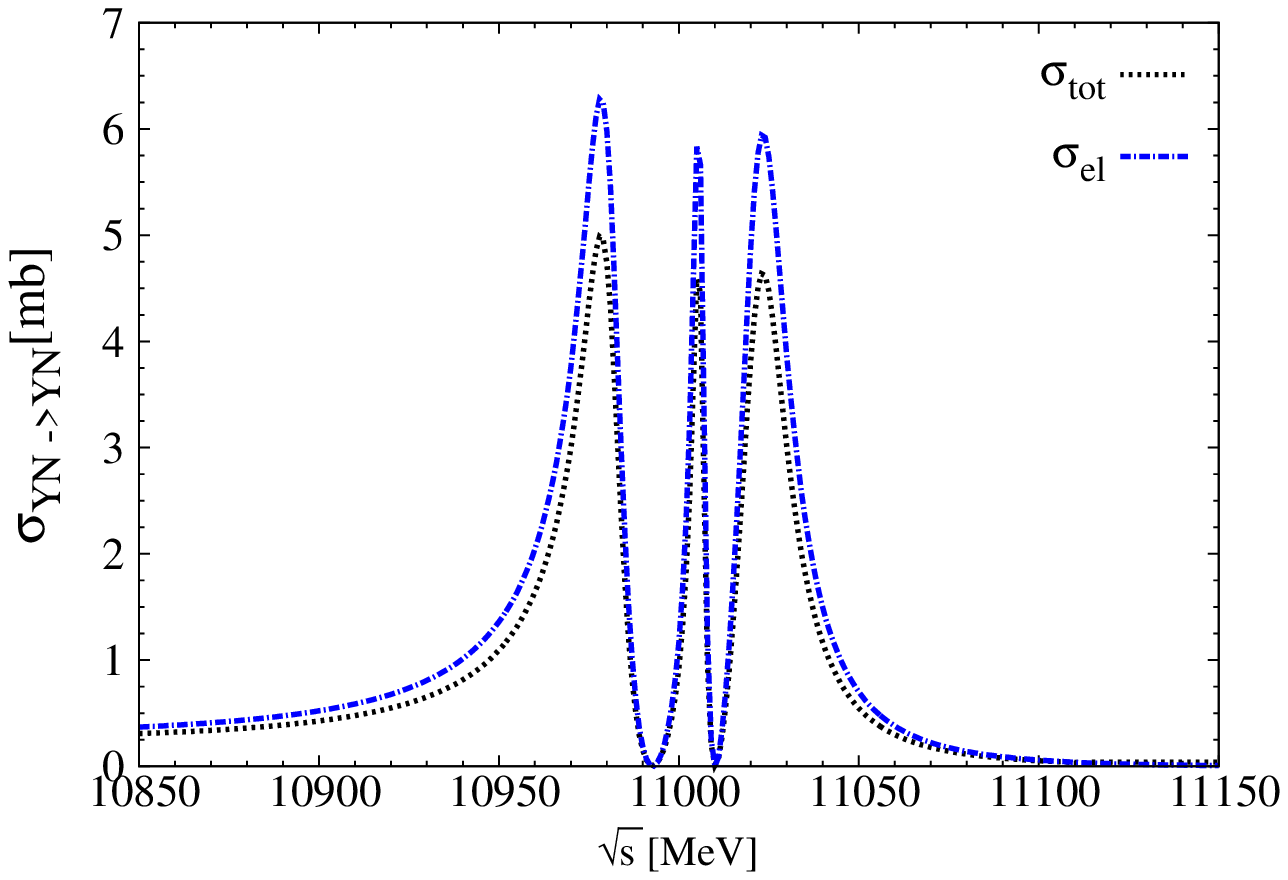}
\caption{Result of the $J=3/2,~I=1/2$ sector with hidden beauty. Left: Modulus squared of the amplitudes. Right: The total, elastic and inelastic cross sections of the $\Upsilon N$ channel. (Analogous to Fig. \ref{fig:charm2})}
\label{fig:beauty2}
\end{figure}

With the heavy quark flavour symmetry and using similar dynamics of the interaction potentials, we extrapolate our formalism into the hidden beauty sector. Our results are shown in Figs.~\ref{fig:beauty1} and \ref{fig:beauty2} for the sectors of $J=1/2,~I=1/2$ and $J=3/2,~I=1/2$ respectively. In the left panel of Fig.~\ref{fig:beauty1}, the predicted states of Ref. \cite{Xiao:2013jla} have been reproduced in the scattering amplitudes of the $B \Sigma_b$, $B^* \Sigma_b$, $B^* \Sigma_b^*$ channels, respectively, where the masses of these three peaks are $10961\mev, ~11002\mev, ~11023\mev$ and the widths are $12\mev, ~26\mev, ~28\mev$, respectively. All of them appear as resonant structures in the total and elastic cross sections of the $\Upsilon N$ channel in the middle panel of Fig.~\ref{fig:beauty1}, where the peak for the third state is nearly washed out. We also find that the second peak is enhanced since it couples to $\Upsilon N$ channel strongly  \cite{Xiao:2013jla}, which leads to an elastic cross section larger than the total one determined by Eq. \eqref{eq:sigtot}. Therefore, in the inelastic cross sections, analogous to the hidden charm sector, the second one disappears and only the other two keep showing up. For the cross sections of the $\eta_b N$ channel, shown in the right panel  of Fig.~\ref{fig:beauty1}, like the results for hidden charm, only the first one and the third one show up in the total and elastic cross sections since the second one of the $B^* \Sigma_b$ couples to the $\eta_b N$ channel weakly \cite{Xiao:2013jla}, and only the first one survives in the inelastic cross sections. Thus, we can see that the $B \Sigma_b$ bound state can be found in both of the $\Upsilon N$ and $\eta_b N$ final state interactions with elastic and inelastic processes. The second one of $B^* \Sigma_b$ only can be seen in the $\Upsilon N$ elastic process. And the state of $B^* \Sigma_b^*$ can be found both elastic and inelastic process of  the $\Upsilon N$ interaction.

In the left panel of Fig.~\ref{fig:beauty2}, one finds three peaks below the thresholds of $B \Sigma_b^*, ~B^* \Sigma_b, ~B^* \Sigma_b^*$, respectively, with the masses $10981\mev, ~11006\mev, ~11022\mev$ and the widths about $14\mev, ~4\mev, ~17\mev$ accordingly, which are consistent with Ref. ~\cite{Xiao:2013jla}. These structures of the resonant peak are also found in the total and elastic cross sections, shown in the right  panel of Fig. \ref{fig:beauty2}, but surprisingly the total cross section determined by the optical theorem, Eq. \eqref{eq:sigtot}, is smaller than the elastic one because of the large momentum transfer in the $\Upsilon N$ channel which reduces the total cross section. Therefore, the three predicted states in Ref. \cite{Xiao:2013jla} can be seen in the elastic $\Upsilon N$ final interactions.

\section{Conclusions}

We have investigated the elastic and inelastic cross sections of the $J/\psi N$, $\eta_c N$, $\Upsilon N$ and $\eta_b N$ channels using an extended local hidden gauge formalism supplemented with constraints from heavy quark spin and flavour symmetry. The predicted bound states in the earlier  works of Refs.~ \cite{Xiao:2013yca,Xiao:2013jla}, $\bar D \Sigma_c$, $\bar D^* \Sigma_c$, $\bar D \Sigma_c^*$, $\bar D^* \Sigma^*_c$ and $B \Sigma_b$, $B^* \Sigma_b$, $B \Sigma_b^*$, $B^* \Sigma^*_b$ with spin $J=1/2, \ 3/2$, should appear  in elastic and/or inelastic final-state interactions of  the $J/\psi N$, $\eta_c N$ and $\Upsilon N$, $\eta_b N$ channels, respectively. 
Clearly, these are only rough estimates due to the various approximations entering the formalism used. Note also that the hidden beauty states have already been discussed in Refs.~\cite{Chen:2015loa,Karliner:2015ina}. Furthermore, since we work in the isospin bases, for experimental searches it is advisable to look for the predicted states with isospin $I=1/2$ in the $J/\psi \ p/n$, $\eta_c \ p/n$, $\Upsilon \ p/n$ and $\eta_b \ p/n$ ($p/n$ are the proton and neutron) elastic or inelastic interaction processes in the future, where the predicted neutral partners $P_c^0$ \cite{Lebed:2015tna} of the $P_c^+$ states \cite{Aaij:2015tga} may be found in these processes.
 
{\it Note added:} Not soon later, Ref. \cite{Karliner:2015voa} also predicts a $B^* \Sigma_b$ state $P_b$, which can be looked for in the $\Upsilon p$ channel.

\section*{Acknowledgments}
We thank C. Hanhart, Q. Wang, J. A. Oller and J. Nieves for useful discussions, and are grateful to E. Oset and F. K. Guo for useful comments and a careful reading the manuscript.
This work is supported in part by the DFG and the NSFC through
funds provided to the Sino-German CRC~110 ``Symmetries and
the Emergence of Structure in QCD''.
This work is also supported in part by The Chinese Academy of Sciences CAS
President's International Fellowship Initiative (PIFI) grant no.
2015VMA076.

\end{document}